\documentclass[prb,preprintnumbers,amsmath,amssymb,floatfix]{revtex4}

\usepackage{graphicx}
\usepackage{subfigure}
\usepackage{dcolumn}

\begin{document}

\title{Quantum theory of light diffraction}
\author{Xiang-Yao Wu$^{a}$ \footnote{E-mail: wuxy2066@163.com}, Bai-Jun Zhang$^{a}$, Jing-Hai Yang$^{a}$, Li-Xin Chi$^{a}$,
Xiao-Jing Liu$^{a}$\\Yi-Heng Wu$^{a}$, Qing-Cai Wang$^{a}$, Yan
Wang$^{a}$, Jing-Wu Li$^{b}$ and Yi-Qing Guo$^{c}$}
\affiliation{a.Institute of Physics, Jilin Normal
University, Siping 136000, China \\
b. Institute of Physics, Xuzhou Normal University, Xuzhou 221000,
China
\\
c. Institute of High Energy Physics, P. O. Box 918(3), Beijing
100049, China }

\begin{abstract}
 At present, the theory of light diffraction only has the simple wave-optical approach.
 In this paper, we study light diffraction with the approach of relativistic quantum theory.
 We find that the slit length, slit width, slit thickness and wave-length of light have
 affected to the diffraction intensity and form of diffraction
 pattern. However, the effect of slit thickness on the diffraction pattern can not
 be explained by wave-optical approach, and it can be explained in quantum
 theory. We compare the theoretical results with single and multiple slits experiment data,
 and find the theoretical results are accordance with the experiment data. Otherwise,
 we give some theory prediction. We think all the new prediction will be tested
  by the light diffraction experiment.

\end{abstract}
\maketitle

{\bf 1. Introduction} \vskip 8pt

It is known that the nonclassical phenomena of two photon
interference [1] and two- photon ghost diffraction and imaging
[2], [3] have classical counterparts. Two photon interference of
classical light has been first discovered in the pioneering
experiments by Hanbury Brown and Twiss [4] and since then was
observed with various sources, including pseudothermal ones [5],
true thermal ones [6], and coherent ones [7]. Somewhat later,
ghost imaging with classical light has been demonstrated, both in
the near-field and far-field domains [8], [9], [10]. The present
optical imaging technologies, such as optical lithography, have
reached a spatial resolution in the sub-micrometer range, which
comes up against the diffraction limit due to the wavelength of
light. However, the guiding principle of such technology is still
based on the classical diffraction theory established by Fresnel,
Kirchhoff and others more than a hundred years ago. Recently, the
use of quantum-correlated photon pairs (biphotons) to overcome the
classical diffraction limit was proposed and attracted much
attention. Obviously, quantum theory approaches are necessary to
explain the diffraction-interference of the quantum-correlated
multi photon state. As is well known, the classical optics with
its standard wave-theoretical methods and approximations, such as
Huygens' and Kirchhoff's theory, has been successfully applied to
classical optics, and has yielded good agreement with many
experiments. However, light interference and diffraction are
quantum phenomena, and its full description needs quantum theory
approach. In 1924, Epstein and Ehrenfest had firstly studied light
diffraction with the old quantum theory, i.e., the quantum
mechanics of correspondence principle, and obtained a identical
result with the classical optics [11]. In this work, we study the
single-slit and multiple-slit diffraction of light with the
approach of relativistic quantum theory of photon. In view of
quantum theory, the light has the nature of wave, and the wave is
described by wave function. As the wave equation which we study
has the character of vector, we choose wave function
$\vec{\psi}(\vec{r},t)$ to describe the wave. The wave function
$\vec{\psi}(\vec{r},t)$ can be calculated with relativistic wave
equation and it also has statistical meaning, i.e.,
$|\vec{\psi}(\vec{r},t)|^{2}$ can be explained as the photon's
probability density at the definite position. In
 light diffraction, because the diffraction intensity $I$ is
directly proportional to  $|\vec{\psi}(\vec{r},t)|^{2}$, we can
obtain the diffraction intensity by calculating the light wave
function $\vec{\psi}(\vec{r},t)$ distributing on display screen,
 and the light wave functions can be
divided into three areas. The first area is the incident area,
where the photon wave function is a plane wave. The second area is
the slit area, where the light wave function can be calculated by
quantum wave equation of light. The third area is the diffraction
area, where the light wave function can be calculated by the
Kirchhoff's law. For multiple-slit diffraction, we can obtain the
total diffraction wave function by superposition the diffraction
wave function of every slit. In the following, we will calculate
these wave functions.

The paper is organized as follows. In section 2 we calculate the
light wave function in the single-slit with quantum theory
approach. In section 3 we calculate the light wave function in
diffraction area with the Kirchhoff's law. Section 4 is
multiple-slit diffraction wave function. Section 5 is numerical
result. Section 6 is a summary of results and conclusion. \vskip
8pt

\setlength{\unitlength}{0.1in}
 \begin{center}
\begin{figure}
\begin{picture}(100,10)
 \put(30,4){\vector(1,0){10}}
 \put(30,4){\vector(0,1){10}}
 \put(30,4){\vector(2,1){8}}
 \put(26,2){\line(1,0){10}}
 \put(26,2){\line(0,1){10}}
 \put(36,2){\line(0,1){10}}
 \put(26,12){\line(1,0){10}}
 \put(30,8){\line(1,0){2}}
 \put(32,4){\line(0,1){4}}
 \put(41,3){\makebox(2,1)[l]{$y$}}
 \put(30,14){\makebox(2,1)[c]{$x$}}
 \put(30,2.6){\makebox(2,1)[l]{$o$}}
 \put(39,8){\makebox(2,1)[c]{$z$}}
 \put(28,6){\makebox(2,1)[c]{$b$}}
 \put(30,8){\makebox(2,1)[c]{$a$}}
\end{picture}
\caption{The single-slit geometry, $a$ is the width and $b$ is the
length of the slit. } \label{moment}
\end{figure}
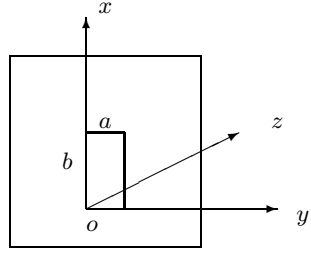
\end{center}

{\bf 2. Quantum approach
 of light single-slit diffraction  }
\vskip 8pt

In an infinite plane, we consider a single-slit, its width $a$ and
length $b$ are shown in FIG. 1. The $x$ axis is along the slit
length and the axis is along the slit width $a$. In the following,
we calculate the light wave function in the single-slit with
relativistic wave equation. At time $t$, we suppose that the
incident plane wave travels along the $z$ axis. It is
\begin{eqnarray}
\vec{\psi}_{0}(z,
t)&=&\vec{A}e^{\frac{i}{\hbar}(pz-Et)}\nonumber\\
&=&\sum_{j}A_{j}\cdot e^{\frac{i}{\hbar}(pz-Et)}\vec{e}_{j}\nonumber\\
&=&\sum_{j}\psi_{0j}\cdot e^{-\frac{i}{\hbar}Et}\vec{e}_{j},
\end{eqnarray}
where $\psi_{0j}=A_{j}\cdot e^{\frac{i}{\hbar}pz}$,  $j= x, y, z$
and $\vec{A}$ is a constant vector. The time-dependent
relativistic wave equation of light is [12]
\begin{equation}
i\hbar\frac{\partial}{\partial
t}\vec{\psi}(\vec{r},t)=c\hbar\nabla\times\vec{\psi}(\vec{r},t)+V\vec{\psi}(\vec{r},t),
\end{equation}
where $c$ is light velocity. From Eq. (2), we can find the light
wave function $\vec{\psi}(\vec{r},t)\rightarrow 0$  when
$V(\vec{r})\rightarrow\infty$. The potential energy of light in
the single-slit is
\begin{eqnarray}
V(x,y,z)&=&0  \hspace{0.3in}0\leq x\leq b, 0\leq y\leq a, 0\leq z\leq c'\nonumber\\
       &=&\infty \hspace{0.3in} otherwise,
\end{eqnarray}
where $c'$ is the slit thickness. We can get the time-dependent
relativistic wave equation in the slit ($V(x,y,z)=0$), it is
\begin{equation}
i\hbar\frac{\partial}{\partial
t}\vec{\psi}(\vec{r},t)=c\hbar\nabla\times\vec{\psi}(\vec{r},t),
\end{equation}
by derivation on  Eq. (4) about the time t and multiplying
$i\hbar$ both sides, we have
\begin{equation}
(i\hbar)^2\frac{\partial^2}{\partial
t^2}\vec{\psi}(\vec{r},t)=c\hbar\nabla\times
i\hbar\frac{\partial}{\partial t}\vec{\psi}(\vec{r},t),
\end{equation}
substituting Eq. (4) into (5), we have
\begin{eqnarray}
\frac{\partial^2}{\partial t^2}\vec{\psi}(\vec{r},t)&
=&-c^2[\nabla(\nabla\cdot\vec{\psi}(\vec{r},t))-\nabla^2\vec{\psi}(\vec{r},t)],
\end{eqnarray}
where the formula $\nabla\times\nabla\times
\vec{B}=\nabla(\nabla\cdot\vec{B})-\nabla^2\vec{B}$. From Ref.
[11], the photon wave function is
$\vec{\psi}(\vec{r},t)=\sqrt{\frac{\varepsilon_{0}}{2}}(\vec{E}(\vec{r},t)+i\sigma
c\vec{B}(\vec{r},t))$, we have
\begin{equation}
\nabla\cdot\vec{\psi}(\vec{r},t)=0,
\end{equation}
from Eq. (6) and (7), we have
\begin{equation}
(\frac{\partial^2}{\partial
t^2}-c^2\nabla^2)\vec{\psi}(\vec{r},t)=0.
\end{equation}

The Eq. (8) is the same as the classical wave equation of light.
Here, it is a quantum wave equation of light, since it is obtained
from the relativistic wave equation (2), and it satisfied the new
quantum boundary condition: when $\vec{\psi}(\vec{r},t)\rightarrow
0$, $V(\vec{r})\rightarrow\infty$. It is different from the
classic boundary condition.

When the photon wave function $\vec{\psi}(\vec{r},t)$ change with
determinate frequency $\omega$, the wave function of photon can be
written as
\begin{equation}
\vec{\psi}(\vec{r},t)=\vec{\psi}(\vec{r})e^{-i\omega t},
\end{equation}
substituting Eq. (9) into (8), we can get
\begin{equation}
\frac{\partial^{2}\vec{\psi}(\vec{r})}{\partial
x^{2}}+\frac{\partial^{2}\vec{\psi}(\vec{r})}{\partial
y^{2}}+\frac{\partial^{2}\vec{\psi}(\vec{r})}{\partial
z^{2}}+\frac{4\pi^{2}}{\\\lambda^{2}}\vec{\psi}(\vec{r})=0,
\end{equation}
and the wave function satisfies boundary conditions
\begin{equation}
\vec{\psi}(0,y,z)=\vec{\psi}(b,y,z)=0,
\end{equation}
\begin{equation}
\vec{\psi}(x,0,z)=\vec{\psi}(x,a,z)=0.
\end{equation}

The photon wave function $\vec{\psi}(\vec{r})$ can be wrote
\begin{eqnarray}
\vec{\psi}(\vec{r})&=&\psi_{x}(\vec{r})\vec{e}_{x}+\psi_{y}(\vec{r})\vec{e}_{y}+\psi_{z}(\vec{r})\vec{e}_{z}\nonumber\\
&=&\sum_{j=x,y,z}\psi_{j}(\vec{r})\vec{e}_{j},
\end{eqnarray}
where $j$ is $x$, $y$ or $z$. Substituting Eq. (13) into (10),
(11) and (12), we have the component equation
\begin{equation}
\frac{\partial^{2}\psi_{j}(\vec{r})}{\partial
x^{2}}+\frac{\partial^{2}\psi_{j}(\vec{r})}{\partial
y^{2}}+\frac{\partial^{2}\psi_{j}(\vec{r})}{\partial
z^{2}}+\frac{4\pi^{2}}{\\\lambda^{2}}\psi_{j}(\vec{r})=0.
\end{equation}
\begin{equation}
\psi_{j}(0,y,z)=\psi_{j}(b,y,z)=0,
\end{equation}
\begin{equation}
\psi_{j}(x,0,z)=\psi_{j}(x,a,z)=0.
\end{equation}

The partial differential equation (14) can be solved by the method
of separation of variable. By writing
\begin{equation}
\psi_{j}(x,y,z)=X_{j}(x)Y_{j}(y)Z_{j}(z).
\end{equation}

From Eq. (14), (15), (16) and (17), we can get the general
solution of Eq. (14)
\begin{equation}
\psi_{j}(x,y,z)=\sum_{mn}D_{mnj}\sin{\frac{n\pi
x}{b}}\sin{\frac{m\pi
y}{a}}e^{i\sqrt{\frac{4\pi^{2}}{\lambda^{2}}-\frac{n^{2}\pi^{2}}{b^{2}}-\frac{m^{2}\pi^{2}}{a^{2}}}z},
\end{equation}
since the wave functions are continuous at $z=0$, we have
\begin{equation}
\vec{\psi}_{0}(x,y,z;t)\mid_{z=0}=\vec{\psi}(x,y,z;t)\mid_{z=0},
\end{equation}
or, equivalently,
\begin{eqnarray}
\psi_{0j}(x,y,z)\mid_{z=0}&=&\psi_{j}(x,y,z)\mid_{z=0}.\hspace{0.3in}(j=x,y,z)
\end{eqnarray}

From Eq. (1), (18) and (20), we obtain the coefficient $D_{mnj}$
by fourier transform
\begin{eqnarray}
D_{mnj}&=&\frac{4}{a
b}\int^{a}_{0}\int^{b}_{0}A_{j}\sin{\frac{n\pi
\xi}{b}}\sin{\frac{m\pi \eta}{a}}d\xi d\eta \nonumber\\
&=&\frac{16A_{j}}{mn\pi^{2}} \hspace{0.6in} m,n, odd  \nonumber\\
&=&0 \hspace{0.9in} otherwise,\hspace{0.4in}(j=x,y,z)
\end{eqnarray}
substituting Eq. (21) into (18), we have
\begin{eqnarray}
\psi_{j}(x,y,z)&=&\sum_{m,n=0}^{\infty}\frac{16A_{j}}{(2m+1)(2n+1)\pi^{2}}\sin{\frac{(2n+1)\pi
x}{b}}\sin{\frac{(2m+1)\pi y}{a}} \nonumber\\&&
e^{i\sqrt{\frac{4\pi^{2}}{\lambda^{2}}-\frac{(2n+1)^{2}\pi^{2}}{b^{2}}
-\frac{(2m+1)^{2}\pi^{2}}{a^{2}}}z}, \hspace{0.6in} (j=x,y,z)
\end{eqnarray}
substituting Eq. (22) into (9) and (13), we can obtain the photon
wave function in slit
\begin{eqnarray}
\vec{\psi}(x,y,z;t)&=&\sum_{j=x,y,z}\psi_{j}(x,y,z,t)\vec{e}_{j}\nonumber\\
&=&\sum_{j=x,y,z}\sum_{m,n=0}^{\infty}\frac{16A_{j}}{(2m+1)(2n+1)\pi^{2}}\sin{\frac{(2n+1)\pi
x}{b}}\sin{\frac{(2m+1)\pi y}{a}} \nonumber\\&&
e^{i\sqrt{\frac{4\pi^{2}}{\lambda^{2}}-\frac{(2n+1)^{2}\pi^{2}}{b^{2}}
-\frac{(2m+1)^{2}\pi^{2}}{a^{2}}}z}e^{-{i}{\omega}t}\vec{e}_{j}.
\end{eqnarray}
We can consider the case of limit, i.e., the slit length $b$ is
infinity, and the Eq. (8) and (10) become
\begin{equation}
\frac{\partial^{2}}{\partial
t^{2}}\vec{\psi}(y,z,t)-c^{2}(\frac{\partial^{2}}{\partial
y^{2}}+\frac{\partial^{2}}{\partial z^{2}})\vec{\psi}(y,z,t)=0,
\end{equation}
\begin{equation}
\frac{\partial^{2}\vec{\psi}(y,z)}{\partial
y^{2}}+\frac{\partial^{2}\vec{\psi}(y,z)}{\partial
z^{2}}+\frac{4\pi^{2}}{\lambda^{2}}\vec{\psi}(y,z)=0,
\end{equation}
we can easily obtain the light wave function in the single-slit
when $b\rightarrow \infty$
\begin{eqnarray}
\vec{\psi}(y,z;t)&=&\sum_{j=y,z}\psi_{j}(x,y,z,t)\vec{e}_{j}\nonumber\\
&=&\sum_{j=x,y,z}\sum_{m=0}^{\infty}\frac{4A_{j}}{(2m+1)\pi}\sin{\frac{(2m+1)\pi
y}{a}} \nonumber\\&& e^{i\sqrt{\frac{4\pi^{2}}{\lambda^{2}}
-\frac{(2m+1)^{2}\pi^{2}}{a^{2}}}z}e^{-{i}{\omega}t}\vec{e}_{j}.
\end{eqnarray}

 {\bf 3. The wave function of light diffraction}
\vskip 8pt In the section 2, we have calculated the photon wave
function in slit. In the following, we will calculate diffraction
wave function. we can calculate the wave function in the
diffraction area. From the slit wave function component
$\psi_{j}(\vec{r},t)$, we can calculate its diffraction wave
function component $\Phi_{j}(\vec{r},t)$ by Kirchhoff's law. It
can be calculated by the formula[13]
\begin{equation}
\Phi_{j}(\vec{r},t)=-\frac{1}{4\pi}\int_{s_{0}}\frac{e^{ikr}}{r}\overrightarrow{n}\cdot[\bigtriangledown^{'}\psi_{j}
+(ik-\frac{1}{r})\frac{\overrightarrow{r}}{r}\psi_{j}]ds.
\end{equation}
the total diffraction wave function is
\begin{eqnarray}
\vec{\Phi}(\vec{r},t)&=&\sum_{j=x,y,z}\Phi_{j}(\vec{r},t)\vec{e}_{j},
\end{eqnarray}
the diffraction area is shown in FIG. 2, where
$k=\frac{2\pi}{\lambda}$ is wave vector, $s_{0}$ is the area of
the single-slit, $\overrightarrow{r}^{'}$ the position of a point
on the surface $(z=c')$, $P$ is an arbitrary point in the
diffraction area, and the $\overrightarrow{n}$ is a unit vector,
which is normal to the surface of the single-slit. From FIG. 2, we
have
\begin{eqnarray}
r&=&R-\frac{\overrightarrow{R}}{R}\cdot\overrightarrow{r}^{'}\nonumber\\
&\approx&R-\frac{\overrightarrow{r}}{r}\cdot\overrightarrow{r}^{'}\nonumber\\
 &=&R-\frac{\overrightarrow{k_{2}}}{k}\cdot\overrightarrow{r}^{'},
\end{eqnarray}
then,

\setlength{\unitlength}{0.1in}
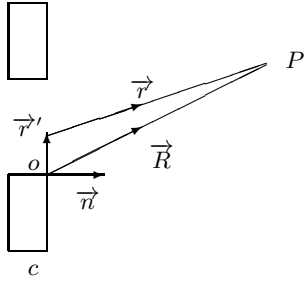
\begin{figure}
\begin{picture}(100,20)
 \put(26,5){\vector(1,0){3}}
 \put(26,5){\vector(0,1){2.2}}
 \put(26,5){\vector(2,1){5}}

 \put(24,1){\line(1,0){2}}
 \put(24,1){\line(0,1){4}}
 \put(26,1){\line(0,1){4}}
 \put(24,5){\line(1,0){2}}

 \put(24,10){\line(1,0){2}}
 \put(24,10){\line(0,1){4}}
 \put(26,10){\line(0,1){4}}
 \put(24,14){\line(1,0){2}}

 \put(26,7){\line(3,1){11.5}}
 \put(26,7){\vector(3,1){5}}
 \put(26,5){\line(2,1){11.5}}
 \put(27.5,3.2){\makebox(2,1)[l]{$\overrightarrow{n}$}}
 \put(24,7){\makebox(2,1)[c]{$\overrightarrow{r}^{\prime}$}}
 \put(25,5){\makebox(2,1)[l]{$o$}}
 \put(31,5.5){\makebox(2,1)[c]{$\overrightarrow{R}$}}
 \put(30,9){\makebox(2,1)[c]{$\overrightarrow{r}$}}
 \put(38.5,10.5){\makebox(2,1)[l]{$P$}}
 \put(25,-0.5){\makebox(2,1)[l]{$c$}}
\end{picture}
\caption{The diffraction area of single-slit} \label{moment}
\end{figure}
\begin{eqnarray}
\frac{e^{ikr}}{r}&=&\frac{e^{ik(R-\frac{\overrightarrow{r}}{r}\cdot\overrightarrow{r}^{'})}}
{R-\frac{\overrightarrow{r}}{r}\cdot\overrightarrow{r}^{'}}\nonumber\\
&\approx&\frac{{e^{ikR}e^{-i\overrightarrow{k_{2}}\cdot\overrightarrow{r}^{'}}}}{R}
\hspace{0.3in}(|\overrightarrow{r}^{'}|\ll R),
\end{eqnarray}
where $\vec{k_{2}}=k\frac{\vec{r}}{r}$. Substituting Eq. (22),
(29) and (30) into (27), one can obtain
\begin{eqnarray}
\Phi_{j}(\vec{r},t)&=&-\frac{e^{ikR}}{4\pi
R}e^{-{i}{\omega}t}\int_{s_{0}}e^{-i\overrightarrow{k_{2}}\cdot
\overrightarrow{r}^{'}}\sum_{m=0}^{\infty}\sum_{n=0}^{\infty}\frac{16A_{j}}{(2m+1)(2n+1)\pi^{2}}\nonumber\\&&
e^{i\sqrt{\frac{4\pi^{2}}{\lambda^{2}}-(\frac{(2n+1)\pi}{b})^{2}-(\frac{(2m+1)\pi}{a})^{2}}\cdot
c'} \sin \frac{(2n+1)\pi}{b}x^{'}\sin
\frac{(2m+1)\pi}{a}y^{'}\nonumber\\&&
[i\sqrt{\frac{4\pi^{2}}{\lambda^{2}}-(\frac{(2n+1)\pi}{b})^{2}-(\frac{(2m+1)\pi}{a})^{2}}+i\overrightarrow{n}\cdot
\overrightarrow{k_{2}}-\frac{\overrightarrow{n}\cdot
\overrightarrow{R}}{R^{2}}]dx^{'}dy^{'}.
\end{eqnarray}
Assume that the angle between $\overrightarrow{k_{2}}$ and $x$
axis ($y$ axis) is $\frac{\pi}{2}-\alpha$ ($\frac{\pi}{2}-\beta$),
and $\alpha (\beta)$ is the angle between $\overrightarrow{k_{2}}$
and the surface of $yz$ ($xz$), then we have
\begin{eqnarray}
k_{2x}=k\sin \alpha,\hspace{0.3in} k_{2y}=k\sin \beta,
\end{eqnarray}
\begin{eqnarray}
\overrightarrow{n}\cdot \overrightarrow{k_{2}}=k\cos \theta,
\end{eqnarray}
where $\theta$ is the angle between $\overrightarrow{k_{2}}$ and
$z$ axis. Substituting Eq. (32) and (33) into (31) gives
\begin{eqnarray}
\Phi_{j}(x,y,z;t)&=&-\frac{e^{ikR}}{4\pi
R}e^{-{i}{\omega}t}\sum_{m=0}^{\infty}\sum_{n=0}^{\infty}\frac{16A_{j}}{(2m+1)(2n+1)\pi^2}
e^{i\sqrt{\frac{4\pi^{2}}{\lambda^{2}}-(\frac{(2n+1)\pi}{b})^{2}-(\frac{(2m+1)\pi}{a})^{2}}\cdot
c'}\nonumber\\&&
[i\sqrt{\frac{4\pi^{2}}{\lambda^{2}}-(\frac{(2n+1)\pi}{b})^{2}-(\frac{(2m+1)\pi}{a})^{2}}+(ik-\frac{1}{R})
\sqrt{\cos^{2}\alpha-\sin^{2}\beta}]\nonumber\\&&
\int^{b}_{0}e^{-ik\sin\alpha\cdot
x^{'}}\sin\frac{(2n+1)\pi}{b}x^{'}dx^{'}\int^{a}_{0}e^{-ik\sin\beta\cdot
y^{'}} \sin \frac{(2m+1)\pi}{a}y^{'}dy^{'}.
\end{eqnarray}
Substituting Eq. (34) into (28), one can get
\begin{eqnarray}
\vec{\Phi}(x,y,z;t)&=&-\frac{e^{ikR}}{4\pi
R}e^{-{i}{\omega}t}\sum_{j=s,y,z}\sum_{m=0}^{\infty}\sum_{n=0}^{\infty}\frac{16A_{j}}{(2m+1)(2n+1)\pi^2}
e^{i\sqrt{\frac{4\pi^{2}}{\lambda^{2}}-(\frac{(2n+1)\pi}{b})^{2}-(\frac{(2m+1)\pi}{a})^{2}}\cdot
c'}\nonumber\\&&
[i\sqrt{\frac{4\pi^{2}}{\lambda^{2}}-(\frac{(2n+1)\pi}{b})^{2}-(\frac{(2m+1)\pi}{a})^{2}}+(ik-\frac{1}{R})
\sqrt{\cos^{2}\alpha-\sin^{2}\beta}]\nonumber\\&&
\int^{b}_{0}e^{-ik\sin\alpha\cdot
x^{'}}\sin\frac{(2n+1)\pi}{b}x^{'}dx^{'}\int^{a}_{0}e^{-ik\sin\beta\cdot
y^{'}} \sin \frac{(2m+1)\pi}{a}y^{'}dy^{'}\vec{e}_{j}.
\end{eqnarray}
Eq. (35) is the  total diffraction wave function in the
diffraction area. From the wave function, we can obtain the
diffraction intensity $I$ on the display screen, we have
\begin{equation}
I\propto|\vec{\Phi}(x,y,z;t)|^{2}.
\end{equation}
\vskip 8pt

{\bf 4. Multiple-slit diffraction wave function of light} \vskip
8pt \setlength{\unitlength}{0.1in}

\hspace{0.2in}From Eq. (23), in the first slit, the photon wave
function $\vec{\psi}_{1}(x,y,z;t)$ is
\begin{eqnarray}
\vec{\psi}_{1}(x,y,z;t)&=&\sum_{j=x,y,z}\sum_{m,n=0}^{\infty}\frac{16A_{j}}{(2m+1)(2n+1)\pi^{2}}\sin{\frac{(2n+1)\pi
x}{b}}\sin{\frac{(2m+1)\pi y}{a}} \nonumber\\&&
e^{i\sqrt{\frac{4\pi^{2}}{\lambda^{2}}-\frac{(2n+1)^{2}\pi^{2}}{b^{2}}
-\frac{(2m+1)^{2}\pi^{2}}{a^{2}}}z}e^{-{i}{\omega}t}\vec{e}_{j}.
\end{eqnarray}

From FIG. 3, in the second slit, when we make the coordinate
translations :
\begin{eqnarray}
&&x'=x\nonumber\\
&&y'=y-(a+d)\nonumber\\
&&z'=z,
\end{eqnarray}
\begin{figure}
\begin{picture}(70,10)
 \put(26,4){\vector(1,0){18}}
 \put(26,4){\vector(0,1){10}}
 \put(26,4){\vector(2,1){15}}
 \put(22,2){\line(1,0){24}}
 \put(22,2){\line(0,1){13}}
 \put(46,2){\line(0,1){13}}
 \put(22,15){\line(1,0){24}}
 \put(26,8){\line(1,0){2}}
 \put(28,4){\line(0,1){4}}
 \put(32,4){\line(0,1){4}}
 \put(32,8){\line(1,0){2}}
 \put(34,4){\line(0,1){4}}
 \put(38,4){\line(0,1){4}}
 \put(40,4){\line(0,1){4}}
 \put(38,8){\line(1,0){2}}
 \put(28,7){\line(1,0){1}}
 \put(31,7){\line(1,0){1}}
 \put(35,5){\makebox(2,1)[l]{$\cdots$}}
 \put(44,2.5){\makebox(2,1)[l]{$y$}}
 \put(26,13){\makebox(2,1)[c]{$x$}}
 \put(25.5,2.6){\makebox(2,1)[l]{$o$}}
 \put(41,11){\makebox(2,1)[c]{$z$}}
 \put(24,6){\makebox(2,1)[c]{$b$}}
 \put(26,8){\makebox(2,1)[c]{$a$}}
 \put(29.5,6.5){\makebox(2,1)[l]{$d$}}
 \put(27,2.6){\makebox(2,1)[l]{$1$}}
 \put(33,2.6){\makebox(2,1)[l]{$2$}}
 \put(39,2.6){\makebox(2,1)[l]{$N$}}
\end{picture}
\caption{Multiple-slit geometry with $a$ the single slit width,
$b$ the slit length and $d$ the distance between the two slit. }
\label{moment}
\end{figure}
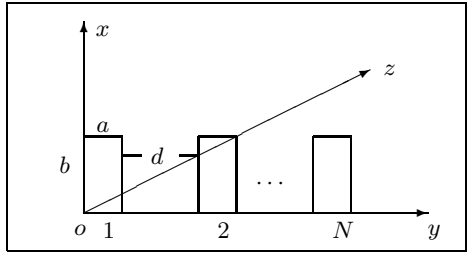
we can obtain the second slit photon wave function by the first
slit photon wave function. It is
\begin{eqnarray}
\vec{\psi}_{2}(x,y,z;t)&=&\sum_{j=x,y,z}\sum_{m,n=0}^{\infty}\frac{16A_{j}}{(2m+1)(2n+1)\pi^{2}}\sin{\frac{(2n+1)\pi
x}{b}}\sin{\frac{(2m+1)\pi [y-(a+d)]}{a}} \nonumber\\&&
e^{i\sqrt{\frac{4\pi^{2}}{\lambda^{2}}-\frac{(2n+1)^{2}\pi^{2}}{b^{2}}
-\frac{(2m+1)^{2}\pi^{2}}{a^{2}}}z}e^{-{i}{\omega}t}\vec{e}_{j}.
\end{eqnarray}
Similarly, we can also obtain the N-th photon wave function. It is
\begin{eqnarray}
\vec{\psi}_{N}(x,y,z;t)&=&\sum_{j=x,y,z}\sum_{m,n=0}^{\infty}\frac{16A_{j}}{(2m+1)(2n+1)\pi^{2}}\sin{\frac{(2n+1)\pi
x}{b}}\sin{\frac{(2m+1)\pi [y-(N-1)(a+d)]}{a}} \nonumber\\&&
e^{i\sqrt{\frac{4\pi^{2}}{\lambda^{2}}-\frac{(2n+1)^{2}\pi^{2}}{b^{2}}
-\frac{(2m+1)^{2}\pi^{2}}{a^{2}}}z}e^{-{i}{\omega}t}\vec{e}_{j}.
\end{eqnarray}
With the Kirchhoff's law, similar as Eq. (35), we can get the
light diffraction wave function in every slit, they are
\begin{eqnarray}
\vec{\Phi}_{1}(x,y,z;t)&=&-\frac{e^{ikR}}{4\pi
R}e^{-{i}{\omega}t}\sum_{j=x,y,z}\sum_{m=0}^{\infty}\sum_{n=0}^{\infty}\frac{16A_{j}}{(2m+1)(2n+1)\pi^2}
e^{i\sqrt{\frac{4\pi^{2}}{\lambda^{2}}-(\frac{(2n+1)\pi}{b})^{2}-(\frac{(2m+1)\pi}{a})^{2}}\cdot
c'}\nonumber\\&&
[i\sqrt{\frac{4\pi^{2}}{\lambda^{2}}-(\frac{(2n+1)\pi}{b})^{2}-(\frac{(2m+1)\pi}{a})^{2}}+(ik-\frac{1}{R})
\sqrt{\cos^{2}\alpha-\sin^{2}\beta}]\nonumber\\&&
\int^{b}_{0}e^{-ik\sin\alpha\cdot
x^{'}}\sin\frac{(2n+1)\pi}{b}x^{'}dx^{'}\int^{a}_{0}e^{-ik\sin\beta\cdot
y^{'}} \sin \frac{(2m+1)\pi}{a}y^{'}dy^{'}\vec{e}_{j}.
\end{eqnarray}
\begin{eqnarray}
\vec{\Phi}_{2}(x,y,z;t)&=&-\frac{e^{ikR}}{4\pi
R}e^{-{i}{\omega}t}\sum_{j=x,y,z}\sum_{m=0}^{\infty}\sum_{n=0}^{\infty}\frac{16A_{j}}{(2m+1)(2n+1)\pi^2}
e^{i\sqrt{\frac{4\pi^{2}}{\lambda^{2}}-(\frac{(2n+1)\pi}{b})^{2}-(\frac{(2m+1)\pi}{a})^{2}}\cdot
c'}\nonumber\\&&
[i\sqrt{\frac{4\pi^{2}}{\lambda^{2}}-(\frac{(2n+1)\pi}{b})^{2}-(\frac{(2m+1)\pi}{a})^{2}}+(ik-\frac{1}{R})
\sqrt{\cos^{2}\alpha-\sin^{2}\beta}]\nonumber\\&&
\int^{b}_{0}e^{-ik\sin\alpha\cdot
x^{'}}\sin\frac{(2n+1)\pi}{b}x^{'}dx^{'}\int^{2a+d}_{a+d}e^{-ik\sin\beta\cdot
y^{'}} \sin \frac{(2m+1)\pi}{a}y^{'}dy^{'}\vec{e}_{j}.
\end{eqnarray}

\begin{eqnarray}
&&......\nonumber\\
&&\nonumber\\
 \vec{\Phi}_{N}(x,y,z;t)&=&-\frac{e^{ikR}}{4\pi
R}e^{-{i}{\omega}t}\sum_{j=x,y,z}\sum_{m=0}^{\infty}\sum_{n=0}^{\infty}\frac{16A_{j}}{(2m+1)(2n+1)\pi^2}
e^{i\sqrt{\frac{4\pi^{2}}{\lambda^{2}}-(\frac{(2n+1)\pi}{b})^{2}-(\frac{(2m+1)\pi}{a})^{2}}\cdot
c'}\nonumber\\&&
[i\sqrt{\frac{4\pi^{2}}{\lambda^{2}}-(\frac{(2n+1)\pi}{b})^{2}-(\frac{(2m+1)\pi}{a})^{2}}+(ik-\frac{1}{R})
\sqrt{\cos^{2}\alpha-\sin^{2}\beta}]\nonumber\\&&
\int^{b}_{0}e^{-ik\sin\alpha\cdot
x^{'}}\sin\frac{(2n+1)\pi}{b}x^{'}dx^{'}\int^{(N-1)(a+d)+a}_{(N-1)(a+d)}e^{-ik\sin\beta\cdot
y^{'}} \sin \frac{(2m+1)\pi}{a}y^{'}dy^{'}\vec{e}_{j}.
\end{eqnarray}
The total diffraction wave function for the N-slit is
\begin{equation}
\vec{\Phi}(x,y,z;t)=\vec{\Phi}_{1}(x,y,z;t)+\vec{\Phi}_{2}(x,y,z;t)+......+\vec{\Phi}_{N}(x,y,z;t).
\end{equation}
From Eq. (44), we can obtain the diffraction intensity $I$ on the
display screen for N-slit, we have
\begin{equation}
I\propto|\vec{\Phi}(x,y,z;t)|^{2}.
\end{equation}
When $b\rightarrow \infty$, we can get the light diffraction wave
function in sing-slit, it is
\begin{eqnarray}
\vec{\Phi}_{b\rightarrow \infty}(y,z;t)&=&-\frac{e^{ikR}}{4\pi
R}e^{-{i}{\omega}t}\sum_{j=x,y,z}\sum_{m=0}^{\infty}\frac{4A_{j}}{(2m+1)\pi}
e^{i\sqrt{\frac{4\pi^{2}}{\lambda^{2}}-(\frac{(2m+1)\pi}{a})^{2}}\cdot
c'}\nonumber\\&&
[i\sqrt{\frac{4\pi^{2}}{\lambda^{2}}-(\frac{(2m+1)\pi}{a})^{2}}+(i
k-\frac{1}{R}) \sqrt{\cos^{2}\alpha-\sin^{2}\beta}]\nonumber\\&&
\int^{a}_{0}e^{-ik\sin\beta\cdot y^{'}} \sin
\frac{(2m+1)\pi}{a}y^{'}dy^{'}\vec{e}_{j}.
\end{eqnarray}
From Eq. (46), we can obtain the diffraction intensity
$I_{b\rightarrow \infty}$ on the display screen for sing-slit when
$b\rightarrow\infty$. It is
\begin{equation}
I_{b\rightarrow \infty}\propto|\vec{\Phi}_{b\rightarrow
\infty}(y,z,t)|^{2}.
\end{equation}
In Ref. [11], the authors had firstly studied light diffraction
with the old quantum theory, i.e., the quantum mechanics of the
correspondence principle. They had considered a light quantum
comes into three dimensional crystal lattice, and calculated the
light momentum loss after collision with quantization condition
and light quantum momentum formula $p=\frac{h}{\lambda}$. They
obtained the relation between the deflecting angle of light
quantum collision with the lattice and the lattice period. By
analyzing, the authors given the expression of electronic density
distributing on grating $\rho=A_{m}sin\frac{2\pi mx}{a}$. The
coefficient $A_{m}$ could be obtained by Fourier analysis. The
diffraction spectrum intensity of the $m^{th}$ order is
proportional to $A_{m}^{2}$. Finally, they had obtained the
intensity formula of diffraction spectrum which was in complete
agreement with the classical diffraction. The formula given a
simple relation between diffraction intensity, slit width,
wavelength of incident light and diffraction angle. In this paper,
we present quantum theory of light diffraction using the framework
of a relativistic quantum theory, and obtain the relation between
diffraction intensity, slit length, slit width, slit thickness,
light wavelength and diffraction angle. By calculating, we can
find the theoretical results are accordance with the experiment
data, and give some new theory prediction.
 \vskip 8pt

{\bf 5. Numerical result} \vskip 8pt

The light diffraction experiment of single and multiple slits had
been reported by H.F.Neiners in 1970 [14]. In experiment [14], the
optical system consists of two convex lens, the focal length $f$,
a diffraction screen of slit length $b$ and slit width $a$ ($b\gg
a$) and a display screen. The laser light source of wave length
$\lambda$ places on the focal plane of the first convex lens, the
first lens makes the light beams parallel incident on the
diffraction screen, the second lens is next to the diffraction
screen, and the display screen places on the focal plane of second
convex lens. In the experiment, the diffraction patterns were
given by photos, and not given experimental data. The author found
his results can be explained excellently by the classical
theoretical formula. It is
\begin{equation}
I=I_{0}\frac{\sin^{2}\beta}{\beta^{2}}\frac{\sin^{2}N\gamma}{\sin^{2}\gamma},
\end{equation}
where
\begin{equation}
\beta=\frac{a\pi\sin\theta}{\lambda},\hspace{0.2in}
\gamma=\frac{(a+d)\pi\sin\theta}{\lambda},
\end{equation}
$\theta$ is diffraction angle, $a$ is the width of slit, $d$ is
the distance from the first slit to the second slit, and $n$ is
the number of slit.

FIG. 4 (a)-(f) show the diffraction patterns from two, three,
 four, five, six and seven slits with
 $\lambda=6.328\times10^{-7}m$, the slit width
 $a=0.88\times10^{-4}m$, $a+d=3.52\times10^{-4}m$ ($d$ is the
 distance between two slits), and the distance between slit and
 display screen $R=4.572m$. In FIG. 4 (a)-(f), the solid curve is
 our theoretical result, and the dot curve is the result of Eq.
 (48), i.e., the diffraction data. In FIG. 4 (a)-(f), we take the same experiment parameters
 in our calculation, and the theoretical input
 parameters are: the slit length $b=3.52\times10^{-4}m$,
and the slit thickness $c'=0.88\times10^{-4}m$.  From FIG. 4 (a),
we can find the calculation result is accordance with experiment
data. Since the ration $\frac{a+d}{a}=4$, we find the orders 4, 8,
12 $\cdots$ are missing. The conclusion is accordance with
classical optics.

FIG. 4 (b) - (f) are multiple slits diffraction patterns
corresponding to slit number N=3, 4, 5, 6, 7. We can find the
calculation results are accordance with experiment data, and there
are $N-2$ secondary maxima and $N-1$ minima between the two
principle maxima. The conclusions are accordance with classical
optics.

FIG. 5 shows the diffraction patterns for single slit. In the
experiment, the light wave length $\lambda=6.328\times10^{-7}m$,
the slit width $a=1.76\times10^{-4}m$, and the distance between
slit and display screen $R=4.572m$. In our calculation, we take
the same experiment parameters above, and the theoretical input
parameters are: the slit length $b=4.0\times10^{-4}m$, the slit
thickness $c'=1.1\times10^{-6}m$, and the diffraction angle
$\alpha=0.001rad$. In FIG. 5, the solid curve is our theoretical
result and the dot curve is the result of Eq. (48), i.e., the
diffraction data. From FIG. 5, we can find the calculation result
shows a good agreement with experiment data. We have compared the
theoretical results with experiment data above. In the following,
we give some theoretical prediction in FIG. 6 - FIG. 12. In
calculation, we take the light wave length
$\lambda=6.328\times10^{-7}m$, the distance $R=4.572m$ and the
diffraction angle $\alpha=0.001rad$.

FIG. 6 is obtained by taking the single slit width as $5a$, $10a$
and $20a$ ($a=1.76\times10^{-4}m$) and $b=4.0\times10^{-4}m$,
$c'=1.1\times10^{-6}m$. From FIG. 6, we can find when the slit
width increases, the diffraction patterns become narrower, and the
diffraction intensity increases.

In FIG. 7, the slit width and slit length are equal ($a=b$), and
it is obtained by taking the slit length and slit width as
$\lambda$, $3\lambda$ and $5\lambda$, and $c'=1.1\times10^{-6}m$.
From FIG. 7, we can find when they increase, the diffraction
patterns become narrower, and the diffraction intensity increase.

In FIG. 8, we can obtain an important result, when
$a=b\leq0.1\lambda$, the total diffraction intensity is zero,
i.e., a very small hole can not produce diffraction phenomenon. It
is because the incident light scatters back completely when the
size of slit is very small.

FIG. 9 is obtained by taking the single slit length as $50b$,
$70b$ and infinity ($b=4.0\times10^{-4}m$) and
$a=1.76\times10^{-4}m$, $c'=1.1\times10^{-6}m$. From FIG. 9, we
can obtain the following conclusions: (1) When the slit length
increases, the diffraction intensity increases. (2) When the slit
length changes, the width of diffraction patterns do not change.

FIG. 10 is obtained by taking the single slit thickness as
$100c'$, $1000c'$, $2000c'$ and $3000c'$ ($c'=1.1\times10^{-6}m$)
and $a=1.76\times10^{-4}m$, $b=4.0\times10^{-4}m$. From FIG. 10,
we can obtain the following conclusions: (1) When the slit
thickness increases, the total diffraction intensity decreases.
(2) When the slit thickness increases, the diffraction patterns
spread over.

FIG. 11 is obtained by taking the single wave length as
$10\lambda$, $20\lambda$ and $50\lambda$
($\lambda=6.328\times10^{-7}m$) and $a=1.76\times10^{-4}m$,
$b=4.0\times10^{-4}m$, $c'=1.1\times10^{-6}m$. From FIG. 11, we
can obtain the following conclusions: (1) When the wave length
decrease, the total diffraction intensity increases and
diffraction patterns become narrow. (2) When the wave length
decrease, the number of diffraction patterns become more.

FIG. 12 is obtained by taking the double slit thickness as $c'$,
$10c'$ and $50c'$ ($c'=0.88\times10^{-4}m$) and
$a=0.88\times10^{-4}m$, $b=4\times0.88\times10^{-4}m$,
$a+d=3.52\times10^{-4}m$. From FIG. 12, we can obtain the
following conclusions: (1) When the slit thickness increases, the
diffraction intensity decreases. (2) In the classical optics, we
know when the ratio $\frac{a+d}{a}=n$ ($n=1,2,3...$), the orders
$n$, $2n$, $3n$, ... are missing in double slit diffraction. We
find when the slit thickness takes $c'$, and the ratio
$\frac{a+d}{a}=4$, the orders 4, 8, 12, ... are missing. (3) When
the slit thickness increases, such as $10c'$ and $50c'$, we find
the missing-order phenomenon disappears.
\\

From FIG. 4 and FIG. 5, we can find our calculation results are
accordance with the experiment data, the classical optic results
and the forepart quantum theory results [13]. From FIG. 6 to FIG.
12, we give some new prediction, which can be tested by light
diffraction experiment.
 \vskip 8pt

{\bf 6. Conclusion} \vskip 8pt

In conclusion, we have studied the single-slit, double-slit and
multiple-slit diffraction of light with the relativistic quantum
mechanical approach. We give the relation among diffraction
intensity, slit length, slit width, slit thickness, wave length of
light and diffraction angle. Our calculation results are
accordance with the experiment data of the single-slit,
double-slit and multiple-slit diffraction. Otherwise, we study the
slit length, slit width, slit thickness, the wave length of light
affect on the diffraction intensity and form of diffraction
pattern. However, the slit length and slit thickness affect on the
diffraction pattern can not be obtained in classical optic. In
double slit diffraction, we find when the ratio $\frac{a+d}{a}=n$
(n=1, 2, 3...), the orders n, 2n, 3n,... are missing. When the
slit thickness increase, the missing order phenomenon disappears.
In multiple slit diffraction, we find there are N-2 secondary
maxima and N-1 minima between the two principle maxima, we think
that all the new prediction in our work can be tested by light
diffraction experiment.
 
\newpage

\begin{figure}[tbp]
\begin{picture}(25,0)
 \put(8,20){\makebox(2,1)[l]}
 \put(19,3.5){\makebox(2,1)[c]}
{\resizebox{7cm}{5.5cm}{\includegraphics{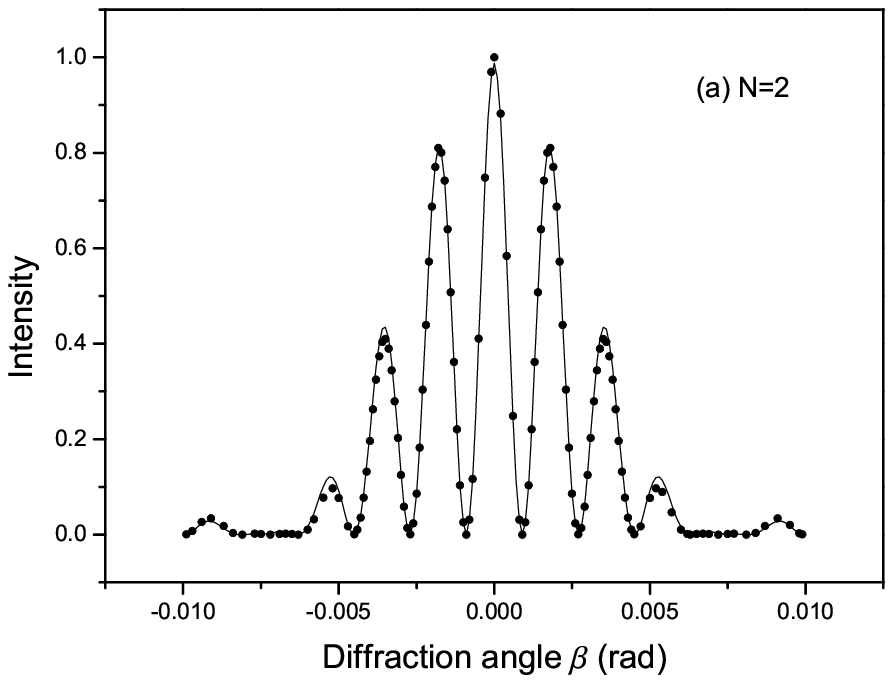}}}
 \put(10,20){\makebox(2,1)[l]}
 \put(20,3.5){\makebox(2,1)[c]}
\end{picture}
{\resizebox{7cm}{5.5cm}{\includegraphics{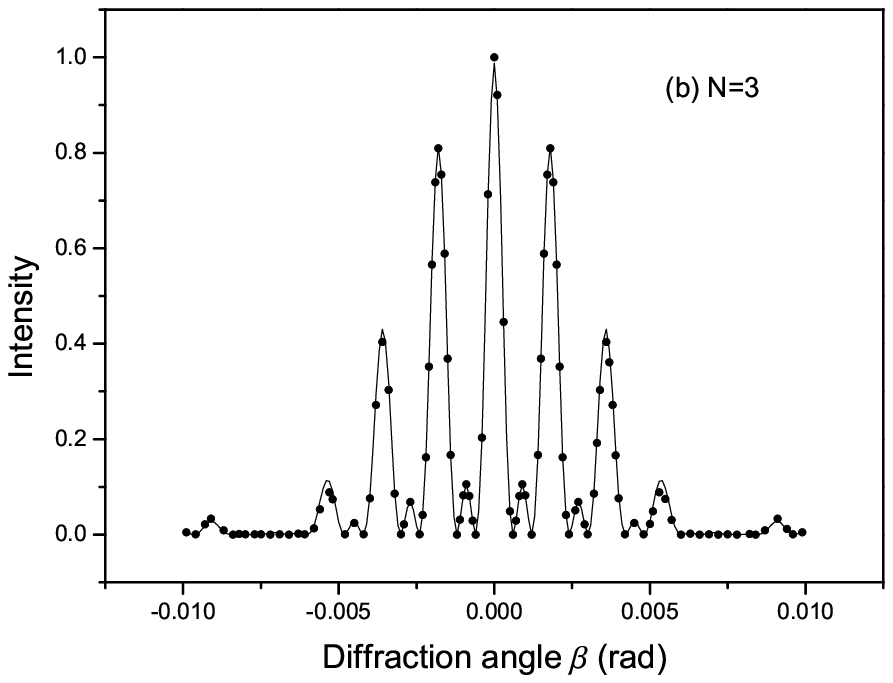}}} \vskip 5pt
\end{figure}

\begin{figure}[tbp]
\begin{picture}(25,0)
 \put(8,20){\makebox(2,1)[l]}
 \put(19,3.5){\makebox(2,1)[c]}
{\resizebox{7cm}{5.5cm}{\includegraphics{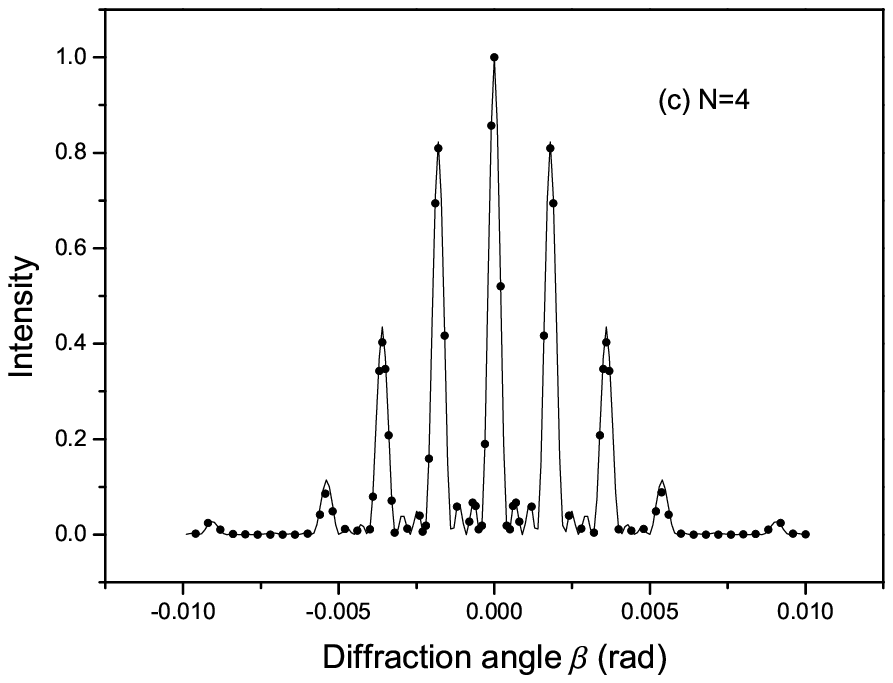}}}
 \put(10,20){\makebox(2,1)[l]}
 \put(20,3.5){\makebox(2,1)[c]}
\end{picture}
{\resizebox{7cm}{5.5cm}{\includegraphics{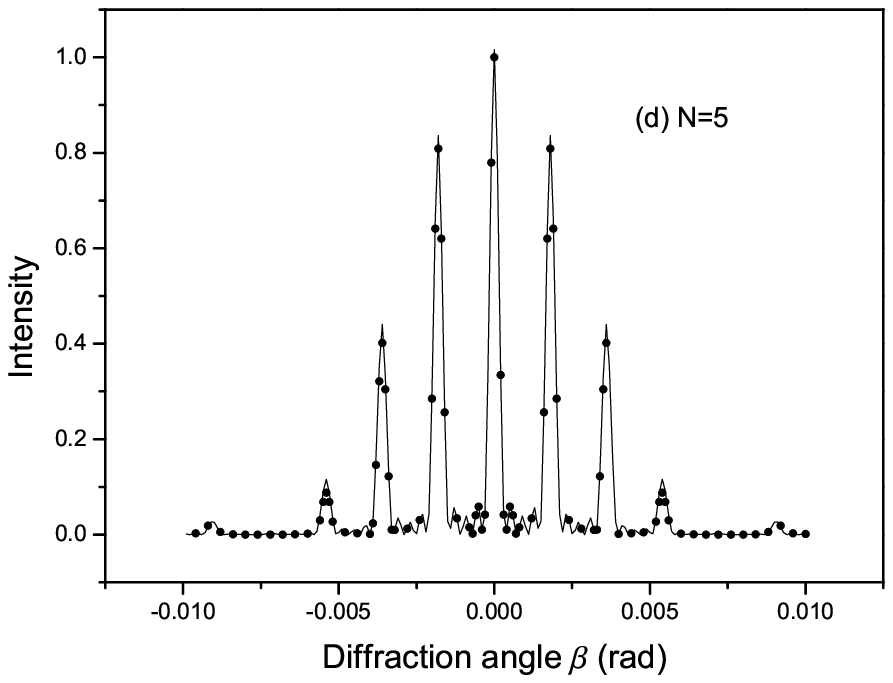}}}\vskip 5pt
\end{figure}

\begin{figure}[tbp]
\begin{picture}(25,0)
 \put(8,20){\makebox(2,1)[l]}
 \put(19,3.5){\makebox(2,1)[c]}
{\resizebox{7cm}{5.5cm}{\includegraphics{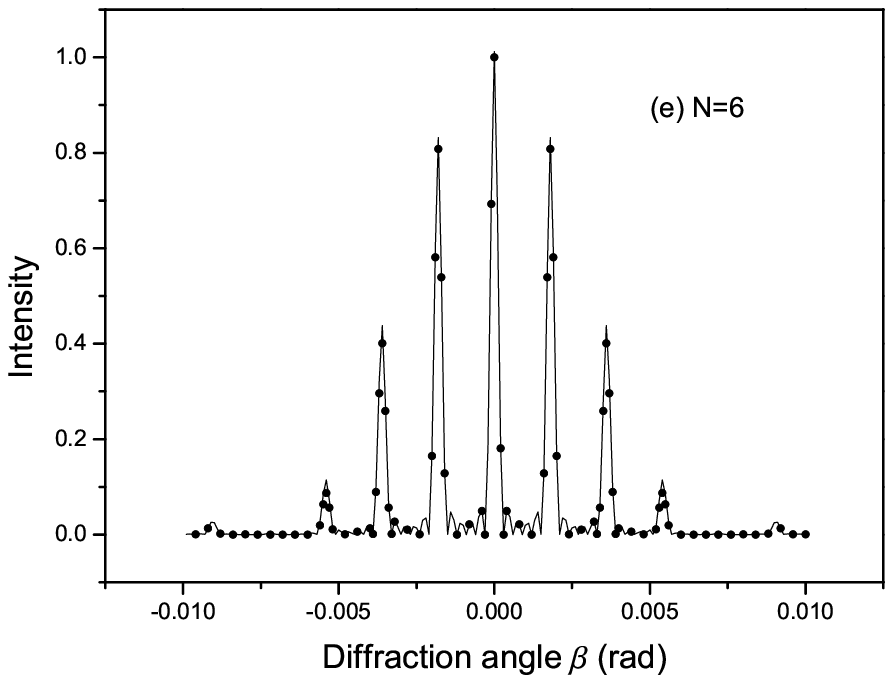}}}
 \put(10,20){\makebox(2,1)[l]}
 \put(20,3.5){\makebox(2,1)[c]}
\end{picture}
{\resizebox{7cm}{5.5cm}{\includegraphics{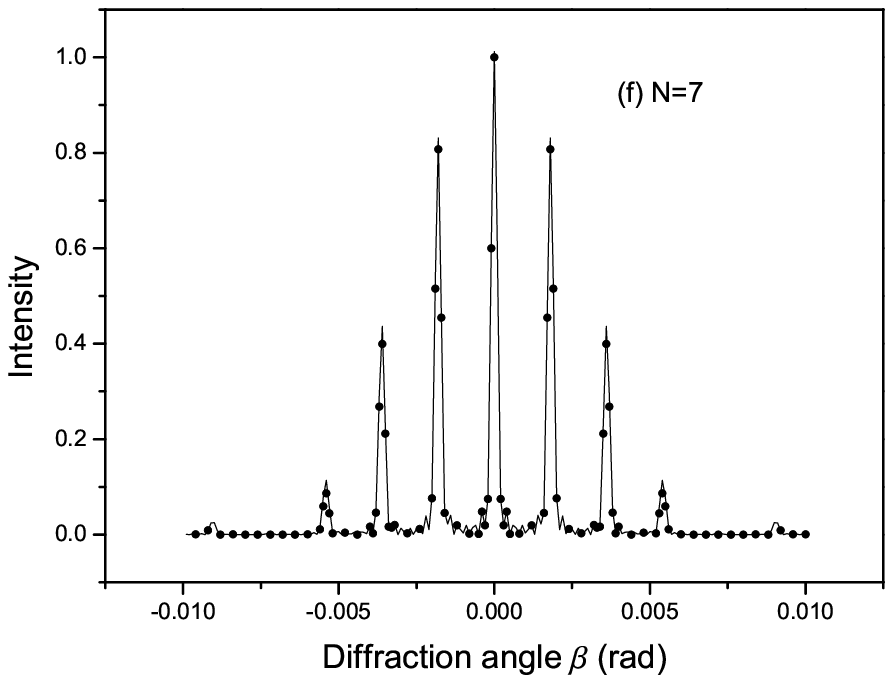}}}\vskip 5pt
{\hspace{0.1in}FIG. 4: The diffraction patterns from two, three,
four, five, six and seven slits with
\\
\hspace{0.4in}$a=0.88\times10^{-4}m$, $a+d=3.52\times10^{-4}m$,
$b=4\times0.88\times10^{-4}m$, $c'=0.88\times10^{-4}m$.\\
\hspace{0.4in}The solid curve is our theoretical calculations and
the dot curve is the result of Eq. (48). }\label{moment}
\end{figure}

\begin{figure}[tbp]
\begin{picture}(25,0)
 \put(8,20){\makebox(2,1)[l]}
 \put(19,3.5){\makebox(2,1)[c]}
{\resizebox{7cm}{5.5cm}{\includegraphics{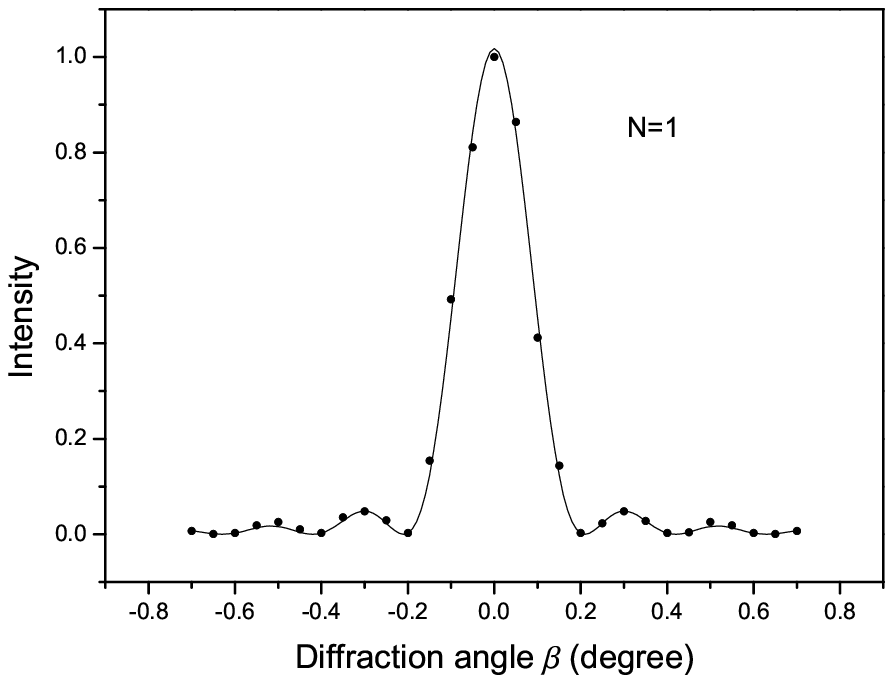}}}
 \put(10,20){\makebox(2,1)[l]}
 \put(20,3.5){\makebox(2,1)[c]}
\end{picture}
{\resizebox{7cm}{5.5cm}{\includegraphics{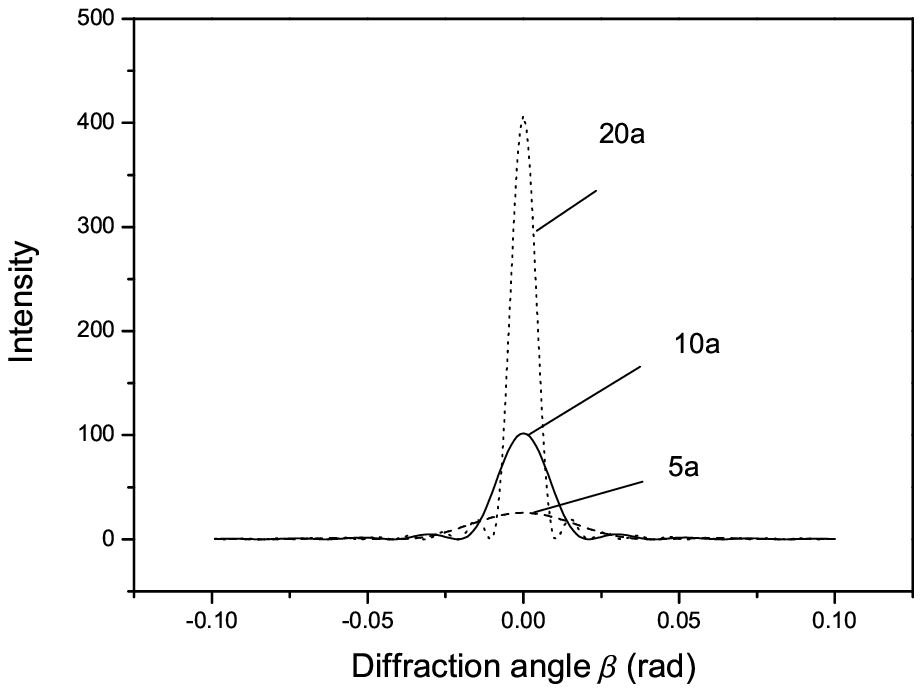}}} \vskip 5pt
{\hspace{0.1in}FIG. 5: The diffraction patterns from single slit
with
\hspace{0.3in} FIG. 6: The diffraction patterns from single slit with\\

\hspace{0.1in} $a=1.76\times10^{-4}m$, $b=4.0\times10^{-4}m$ and
$c'=1.1\times10^{-6}m$. \hspace{0.2in} $b=4.0\times10^{-4}m$ and
$c'=1.1\times10^{-6}m$. The dash,\\

\hspace{0.1in} The solid curve is our theoretical results and the
dot curve \hspace{0.3in}solid and dot curves correspond to slit
width $5a$,

\hspace{0.1in}is the result of Eq. (48). \hspace{2in}  $10a$ and
$20a$($a=1.76\times10^{-4}m$), respectively.}
 \label{moment}
\end{figure}

\begin{figure}[tbp]
\begin{picture}(25,0)
 \put(8,20){\makebox(2,1)[l]}
 \put(19,3.5){\makebox(2,1)[c]}
{\resizebox{7cm}{5.5cm}{\includegraphics{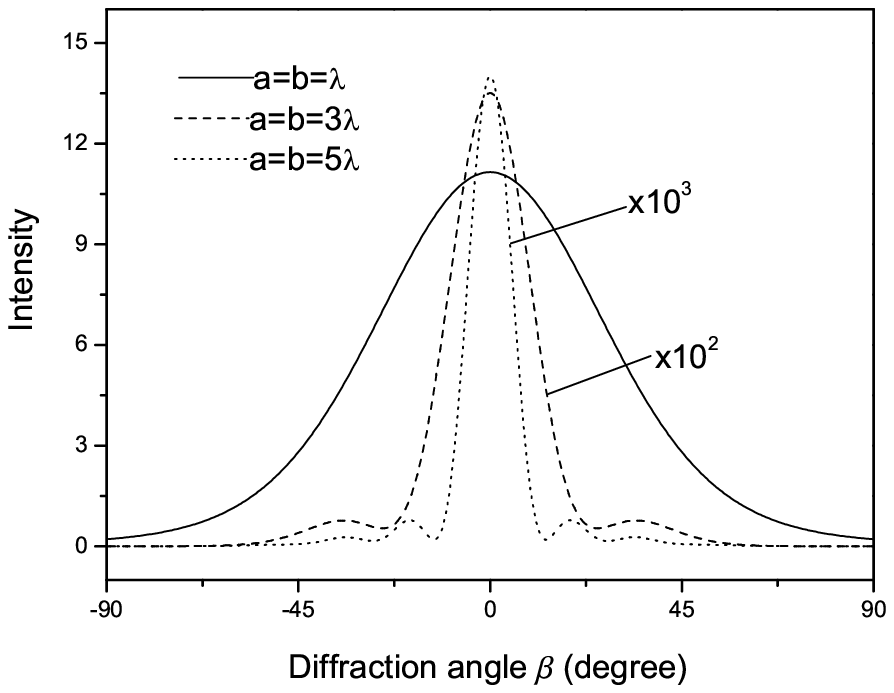}}}
 \put(10,20){\makebox(2,1)[l]}
 \put(20,3.5){\makebox(2,1)[c]}
\end{picture}
{\resizebox{7cm}{5.5cm}{\includegraphics{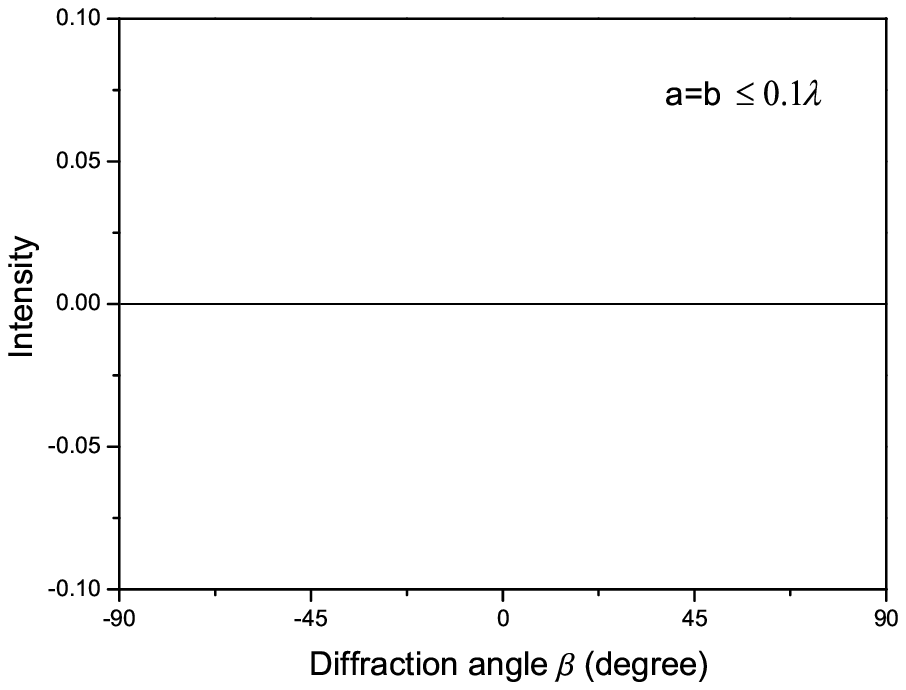}}} \vskip 5pt
{\hspace{0.1in}FIG. 7: The diffraction patterns from single slit
with
\hspace{0.3in} FIG. 8: The diffraction patterns from single slit \\

\hspace{0.1in} $c'=1.1\times10^{-6}m$. The solid, dash and dot
curves  \hspace{0.4in}
 with $c'=1.1\times10^{-6}m$ and $a=b\leq 0.1\lambda$.

\hspace{0.15in} correspond to $a=b=\lambda$, $a=b=3\lambda$ and
$a=b=5\lambda$. \hspace{3in}

\hspace{0.12in} Their real intensity $I$ should be multiplied by
$10^{2}$ and $10^{3}$. \hspace{2.8in}}
 \label{moment}

\end{figure}

\begin{figure}[tbp]
\begin{picture}(25,0)
 \put(8,20){\makebox(2,1)[l]}
 \put(19,3.5){\makebox(2,1)[c]}
{\resizebox{7cm}{5.5cm}{\includegraphics{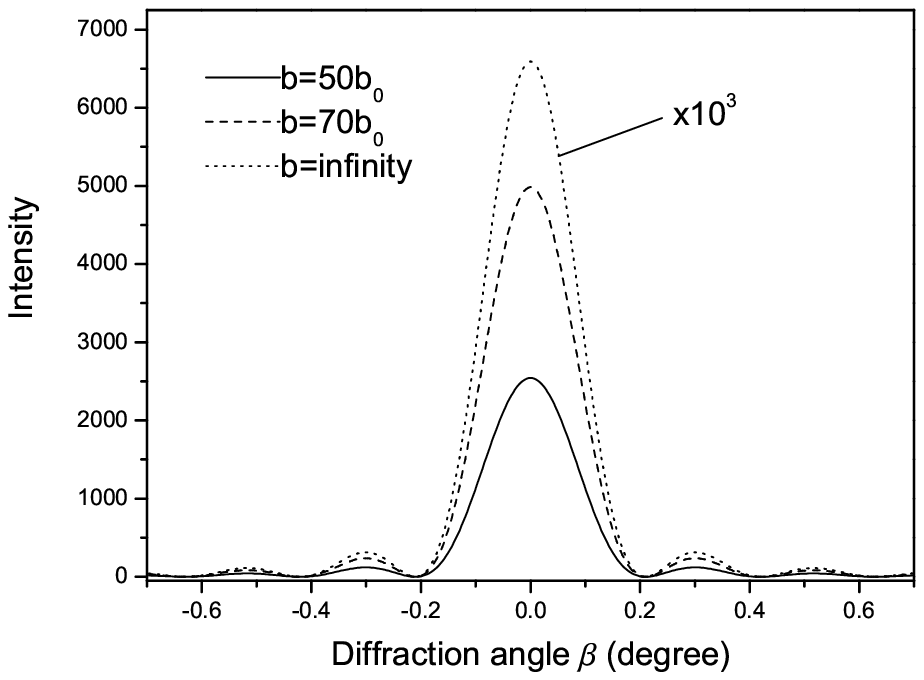}}}
 \put(10,20){\makebox(2,1)[l]}
 \put(20,3.5){\makebox(2,1)[c]}
\end{picture}
{\resizebox{7cm}{5.5cm}{\includegraphics{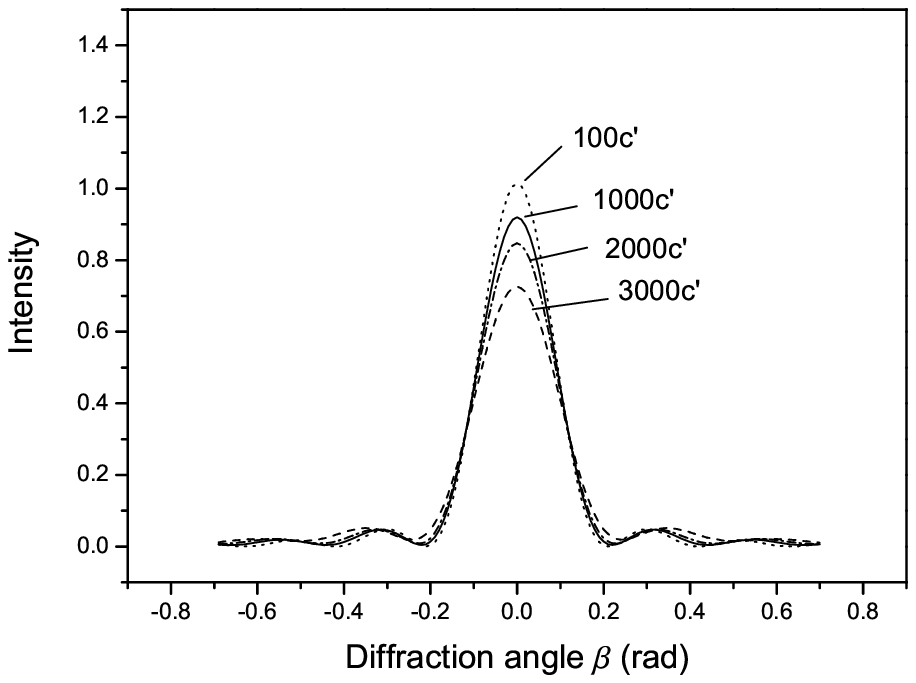}}} \vskip 5pt
{\hspace{0.1in}FIG. 9: The diffraction patterns from single slit
with
\hspace{0.3in} FIG. 10:The diffraction patterns from single slit with \\

 \hspace{0.1in} $a=1.76\times10^{-4}m$ and
 $c'=1.1\times10^{-6}m$. The solid,
 \hspace{0.3in}$a=1.76\times10^{-4}m$ and $b=4.0\times10^{-4}m$. The dot, solid, \\

\hspace{0.05in} dash and dot curves curves correspond to slit
length  \hspace{0.1in} dash-dot and dash curves correspond
to slit thickness $100c'$,\\

\hspace{0.01in} $50b_{0}$, $70b_{0}$ and infinity
($b_{0}=4.0\times10^{-4}m$), respectively. \hspace{0.01in}
$1000c'$, $2000c'$ and $3000c'$ ($c'=1.1\times10^{-6}m$),
respectively.

\hspace{0.05in} The real intensity $I$ should be multiplied by
$10^{3}$. \hspace{3.5in}}
 \label{moment}

\end{figure}

\begin{figure}[tbp]
\begin{picture}(25,0)
 \put(8,20){\makebox(2,1)[l]}
 \put(19,3.5){\makebox(2,1)[c]}
{\resizebox{7cm}{5.5cm}{\includegraphics{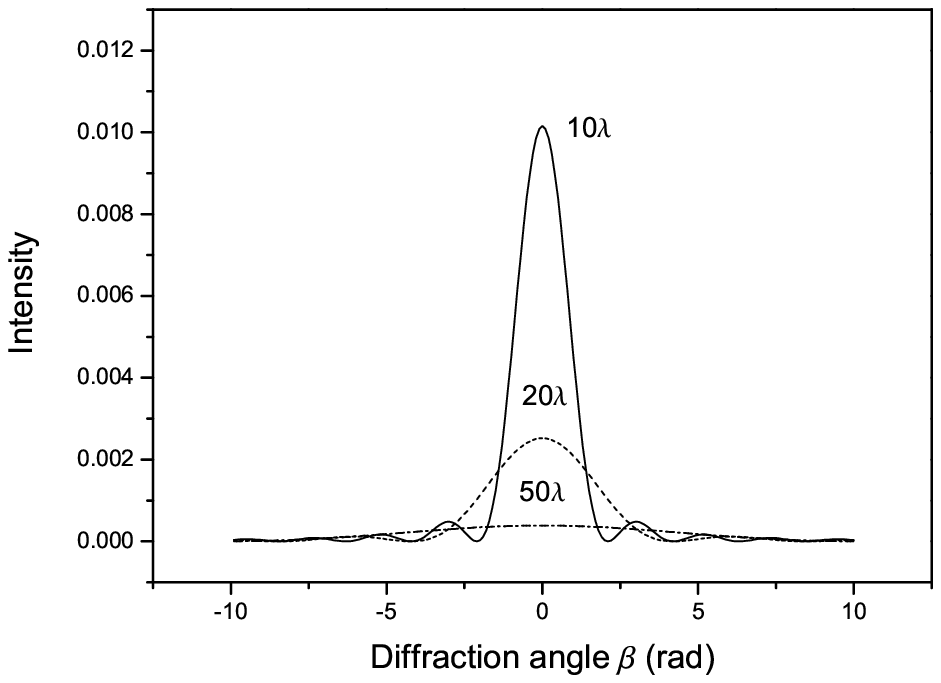}}}
 \put(10,20){\makebox(2,1)[l]}
 \put(20,3.5){\makebox(2,1)[c]}
\end{picture}
{\resizebox{7cm}{5.5cm}{\includegraphics{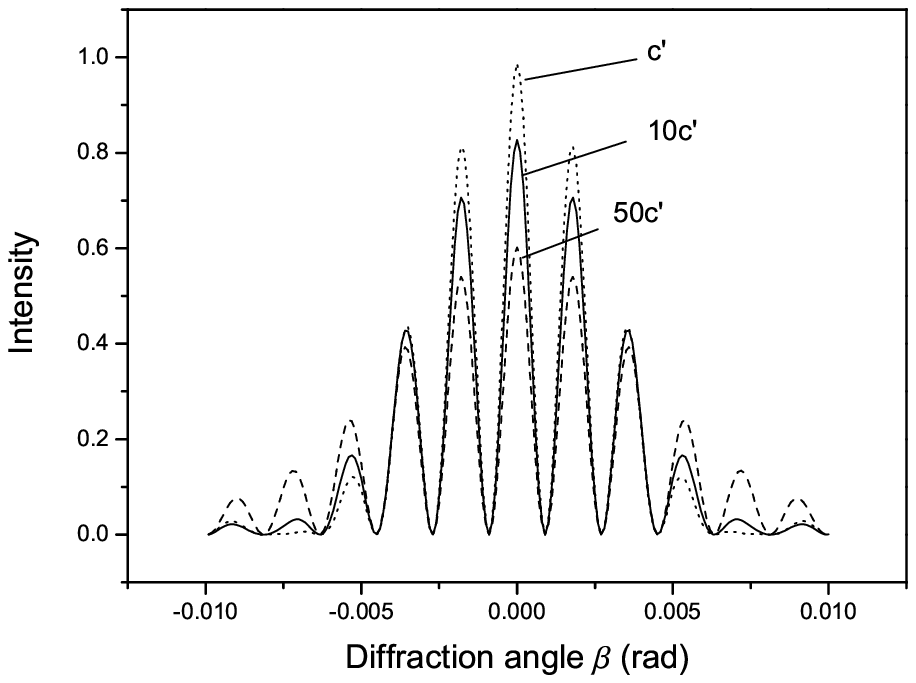}}} \vskip 5pt
{\hspace{0.1in}FIG. 11: The diffraction patterns from single slit
\hspace{0.2in} FIG. 12:The diffraction patterns from double slit \\

 \hspace{0.1in}with $a=1.76\times10^{-4}m$, $b=4.0\times10^{-4}m$ and
 \hspace{0.4in}with $a=0.88\times10^{-4}m$, $b=4\times0.88\times10^{-4}m$ and \\

\hspace{0.05in}$c'=1.1\times10^{-6}m$. The solid, dash and
dash-dot  \hspace{0.3in} $a+d=3.52\times10^{-4}m$. The dot, solid
and dash  \\

\hspace{0.05in} curves correspond to wave length $10\lambda$,
$20\lambda$ and $50\lambda$. \hspace{0.3in} curves correspond to
slit thickness $c'$, $10c'$ and $50c'$.}
 \label{moment}
\end{figure}

\end{document}